\documentstyle[emulateapj,epsfig]{article}

\begin{document}

\title{Deeply penetrating banded zonal flows in the solar 
convection zone}

\author{R. Howe$^1$, J. Christensen-Dalsgaard$^2$, F. Hill$^1$, 
R.W. Komm$^1$, R.M. Larsen$^3$,\\
J. Schou$^3$,  M.J. Thompson$^4$, and J. Toomre$^5$}

\affil{$^1$National Solar Observatory, 950 N. Cherry Avenue, Tucson AZ
85726-6732, USA}
\affil{$^2$Teoretisk Astrofysik Center, Danmarks Grundforskningfond; and\\
Institut for Fysik og Astronomi, Aarhus Universitet, DK-8000 Aarhus C, 
Denmark}
\affil{$^3$HEPL Annex A201, Stanford University, Stanford CA 94305-4085, 
USA}
\affil{$^4$Astronomy Unit, Queen Mary \& Westfield College, London E1 4NS, 
UK}
\affil{$^5$JILA, and Dept. of Astrophysical and Planetary Sciences,
University of Colorado,\\
Boulder CO 80309-0440, USA}

\begin{abstract}

Helioseismic observations have detected small temporal variations of
the rotation rate below the solar surface corresponding to the
so-called `torsional oscillations' known from Doppler measurements of
the surface.  These appear as bands of slower and faster than average
rotation moving equatorward.  Here we establish, using complementary
helioseismic observations over four years from the GONG network and
from the MDI instrument on board SOHO, that the banded flows are not
merely a near-surface phenomenon:  rather they extend downward at least
60 Mm (some 8\% of the total solar radius) and thus are evident over a
significant fraction of the nearly 200 Mm depth of the solar convection
zone.

\end{abstract}



\section{Introduction}

The intensely turbulent state of the solar convection zone is revealed
by the patterns of granulation, mesogranulation and supergranulation
evident in its surface layers (e.g., Brummell, Cattaneo \& Toomre
1995).  Yet accompanying such turbulent and seemingly chaotic
small-scale dynamics are also signs of ordered large-scale behavior.
Most notably the solar differential rotation involves a relatively
smooth decrease in angular velocity from equator to pole, both in the
surface layers (e.g., Snodgrass 1984) and within the convection zone as
inferred from helioseismic measurements (e.g., Thompson et al.  1996;
Schou et al.  1998a).  On the largest scales, the magnetic activity
similarly exhibits well-defined rules as the 22-year cycle progresses.

An enticing link between the latitudes of field emergence and small
variations in the rotation rate of the surface layers is provided by
bands of slightly faster and slower than average zonal flows, called
torsional oscillations, that were observed from direct Doppler
measurements to migrate towards the equator in a manner similar to the
zones of solar activity (e.g., Howard \& LaBonte 1980; Snodgrass,
Howard \& Webster 1985; Ulrich 1998).  Helioseismic analysis of data
from the Michelson Doppler Imager (MDI) instrument (e.g., Scherrer et
al. 1995) on the Solar and Heliospheric Observatory (SOHO) spacecraft
has confirmed the presence of such bands of weak zonal flow, and their
drift towards the equator, for the present solar cycle (Kosovichev \&
Schou 1997; Schou et al. 1998a,b; Schou 1999).  Although the causal
relation between these banded flows and the zones of magnetic activity
is still unclear, it is important to understand whether the flows are
confined to the layer of rotational shear just below the solar
surface.  In this letter, we address such questions using two extensive
helioseismic data sets, covering slightly over four years, obtained
with MDI and with the ground-based Global Oscillation Network Group
(GONG) project (e.g., Harvey et al. 1996).  We establish the
consistency of the independent determinations of the flow from the two
data sets, and infer that the zonal banding signature extends to depths
of about 60 Mm (or about 8\% in radius) below the solar surface.  Thus
these are not superficial features, and provide evidence of ordered
rotational responses as the magnetic cycle is progressing.  More
extensive accounts of such analyses of zonal flows are provided for
GONG data by Howe, Komm \& Hill (2000) and for MDI data by Toomre et al.
(2000).

\vskip-20truept

\section{Observations and data analyses}

The rotation rate of the solar convection zone has been inferred
through inversion of observed rotational splittings of solar f and p
modes.  Two sets of observations have been used.  One was obtained by
the GONG network over the period 1995 May 7 to 1999 June 26.  This set
consists of 40 overlapping series of 108 days each, with starting dates
36 days apart.  The second set was obtained by MDI and consists of 11
contiguous sets from 1996 May 1 to 1998 June 24 (before control of the
spacecraft was lost), one set from 1998 October 23 to 1998 December 21
(after control was reasserted), and four sets from 1999 February 3 to
1999 November 17.  While each of the sets nominally covers 72 days, the
ones surrounding the loss of contact are missing a few days.

%
\begin{figure*}[t]
\centerline{\epsfysize=5.0cm \epsfbox{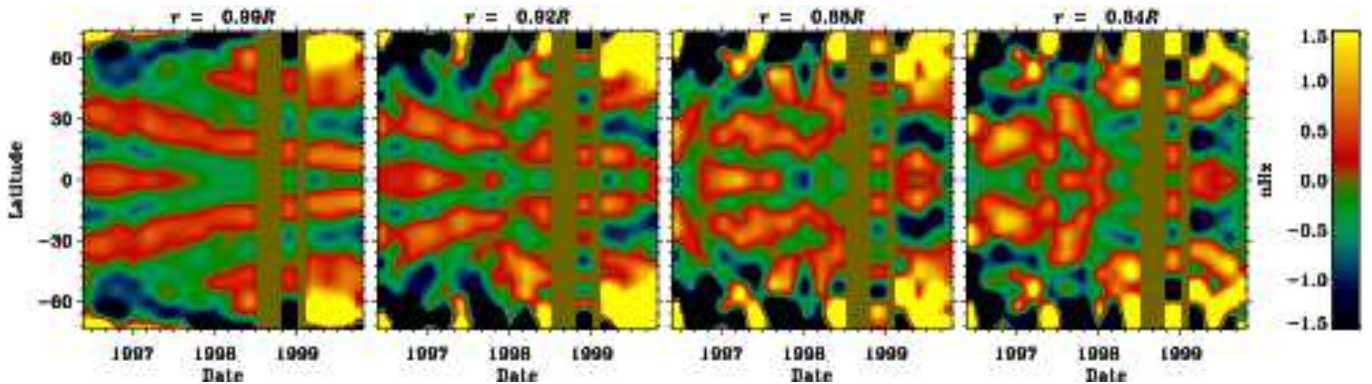}}
\figcaption{
Variation of rotation rate with latitude and time from which a temporal
average has been subtracted to reveal the migrating banded zonal flows,
based on OLA inversion of MDI data for target radii of $0.99 R$, $0.92
R$, $0.88 R$ and $0.84 R$.  The rendition is smoothed in the temporal
direction over a window of 72 days, equal to the period of each
separate set of observations. Uniform olive-green vertical bands
indicate that no data were available in that time period for our
analysis.  The color bar indicates the dynamic range in nHz of the
angular velocity.  
\label{fig:F1}
}
\vspace{-3mm}
\end{figure*}

\placefigure{fig:F1}

The dependence of the frequencies on the azimuthal order was
represented in terms of an expansion on orthogonal polynomials
(Ritzwoller \& Lavely 1991), expressed in terms of the so-called $a$
coefficients $a_k(n,l)$, depending on the radial order $n$ and the
degree $l$ of the mode, as well as on the order $k$ of the
coefficient.  The odd coefficients are related to the angular velocity
$\Omega(r, \theta)$ (as a function of the distance $r$ to the solar
center and the co-latitude $\theta$) by
\begin{equation}
2 \pi~ a_{2 s +1} (n,l)
= \int_0^R \int_0^\pi K_{nls}^{(a)}(r, \theta)~ \Omega(r, \theta)~ 
r {\rm d} r {\rm d} \theta \; ,
\end{equation}
where $R$ is the solar radius and the kernels $K_{nls}^{(a)}$ are
assumed known from a solar model.  The GONG data comprised around
10,000 coefficients, up to $a_{15}$, for a total of typically 1,200
p-mode multiplets $(n,l)$ for $l \le 150$, whereas the MDI sets
contained approximately 30,000 coefficients, up to $a_{35}$, for
roughly 1,800 multiplets with $l \le 300$.  The inversions of the
relations (1) were carried out by means of two methods, described by
Schou et al. (1998a):  two-dimensional regularized least-squares
fitting (RLS) and two-dimensional subtractive optimally localized
averages (OLA).

\vskip-20truept
\section{Results and discussion}

The overall features of the inferred rotation profile are very similar
to those obtained by Schou et al. (1998a) based on 144 days of MDI
data.  Here we concentrate on the time-dependent aspects of the
dynamics of the upper portions of the convection zone.  These are most
readily studied by considering departures of the reconstructed rotation
rate from its temporal average $\bar \Omega(r, \theta)$.
Figure~\ref{fig:F1} shows the evolution of these residuals as a
function of latitude at four target depths, using OLA inversion of the
MDI data.  The inversion is only sensitive to the component of rotation
symmetric around the equator; even so, to show more clearly the
evolution of the features, we have included both hemispheres in the
plots.  The residuals show alternating bands of positive and negative
zonal velocity, relative to the temporal mean, converging towards the
equator with increasing time.  The amplitude in these flows is around
1.5~nHz, corresponding to velocities of up to around $6 \, {\rm m} \,
{\rm s}^{-1}$.  The flows are visible, at roughly the same amplitude,
in the inversion targeted at $0.92 R$, and faint traces are visible in
the inversion targeted at $0.88R$; we return to the significance of
this below.

Although the signal shown in Figure~\ref{fig:F1} seems strong and
coherent, some doubt about its reality may remain.  Thus access to the
independent GONG dataset is essential, with the additional advantage
that it starts almost a year before the MDI data.  As a further test,
we also apply two different analysis techniques to the set, as shown in
Figure~\ref{fig:F2}.  The GONG and MDI data, where they overlap in
time, are essentially consistent at $r= 0.99 R$.  At $r = 0.95 R$ the
GONG RLS reconstructions are somewhat noisier so that the subtle
signature of the migrating zonal bands is less obvious.  The RLS and
OLA inversion results for the MDI data agree very well at both depths
illustrated.

To provide a more quantitative comparison, Figure~\ref{fig:F3} shows
the GONG and MDI solutions at selected radii and latitudes as a
function of time.  It is evident that the three sets of results largely
agree within their error bars.  Also, the variations are highly
statistically significant: at the equator the angular velocity
decreases uniformly, whereas at latitude $30^\circ$, for example, the
variation reflects mainly the passage of the band of more rapid
rotation.  The inversions reveal a considerably different behavior at
the two depths at latitude $60^\circ$, much as could also be inferred
from Figure~\ref{fig:F2}.

\placefigure{fig:F2}

The interpretation of any inversion results must take into account
their finite resolution, as well as the properties of errors of the
inferences.  In particular, the solution obtained at a given location
contains contributions from $\Omega$ at other points, while error
correlation between the solutions at different locations may give the
impression of coherent structure where none exists.  Quantitative
measures of these effects can be obtained from detailed analyses of the
inversion (e.g. Schou et al. 1998a; Howe \& Thompson 1996).  As an
alternative, we have considered artificial data for a number of
prescribed rotation laws, with error properties corresponding to those
of the solar data.  The mode set and errors used correspond to those of
a typical MDI set.  Superimposed on a smoothly varying flow in the
latitudinal direction, the artificial rotation profiles possess a
single pair of flows, $10^\circ$ wide, and rotating 3 nHz faster than
the background level, moving equatorwards at $5^\circ$ per sample.
There are nine samples in all in each `time' sequence.  Three cases are
illustrated in Figure~\ref{fig:F4} using OLA inversions: in the first,
the flows extend from the surface to a depth of 5\% of the solar
radius, in the second to a depth of 8\% and in the third to a depth of
20\%.  For all cases, the branches of the flow are strongly visible at
0.99$R$, and the flows are evident to about the depth to which they are
imposed, while disappearing below that depth.  The 0.92$R$ case most
resembles the solar observations illustrated in Figure~\ref{fig:F1},
whereas the disappearance of the flows in the 0.95$R$ case at 0.93$R$
and the visibility of the 0.80$R$ flow at 0.84$R$ are both inconsistent
with the solar observations.  This evidence, taken together with many
other cases which we have tested, strongly suggests that the solar flow
structure extends to a depth of at least 0.08$R$ with substantial
amplitude, but does not extend much further than 0.10$R$.  The
latitudinal width of the Sun's banded flows (Fig.~\ref{fig:F1}) is
similar to the $10^\circ$ width assumed in the artificial data: this is
consistent also with the Doppler measurements at the surface (Ulrich,
1998). Comparison of Figures \ref{fig:F1} and \ref{fig:F4} suggests
that the solar flows may be somewhat weaker than the $3\,{\rm nHz}$
chosen for the artificial case.

%
{\center\epsfig{file=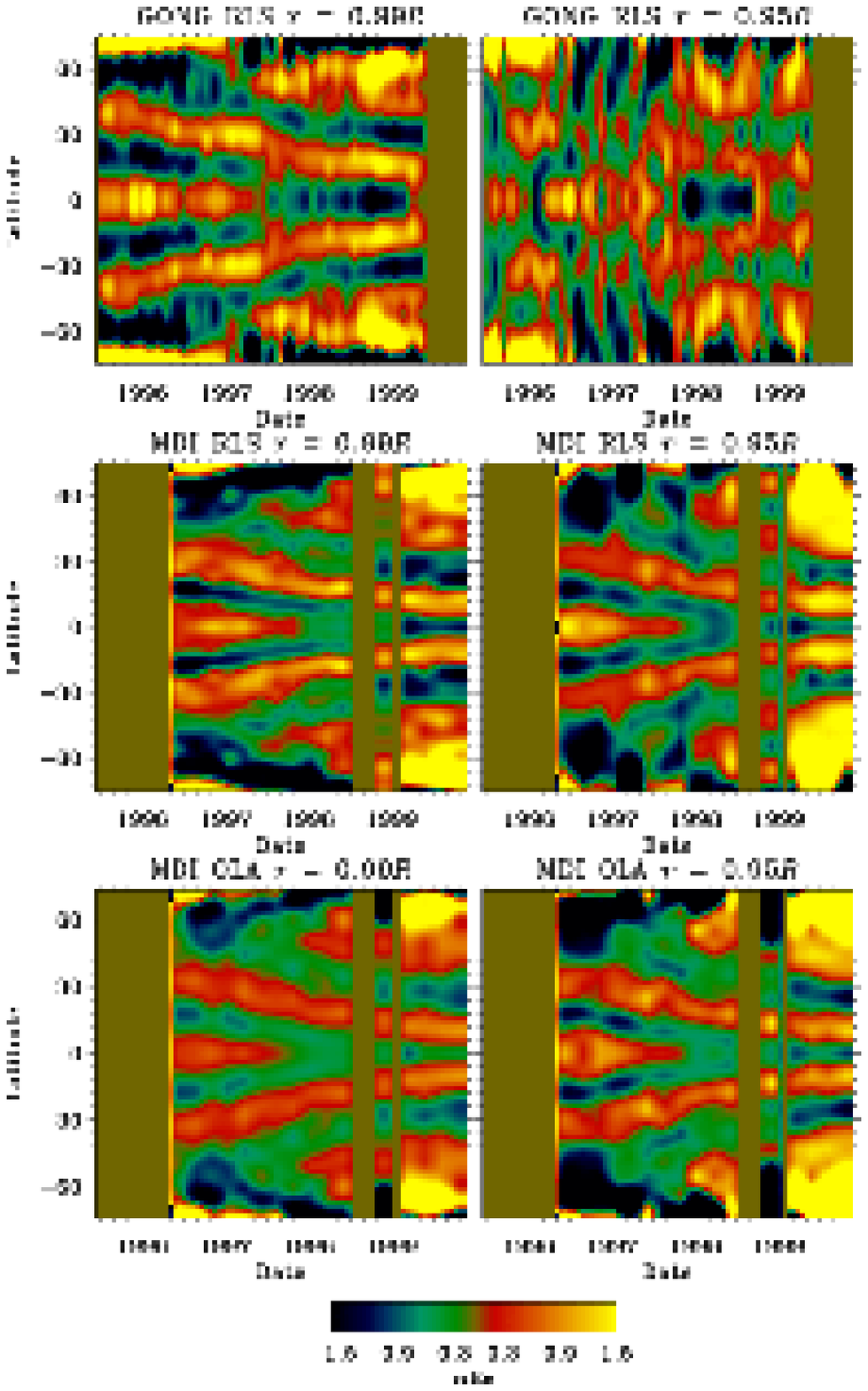,width=\linewidth}}
\figcaption{
Comparison of various inversions of GONG and MDI data showing evolution
of residual rotation rate (cf. Fig.~\ref{fig:F1}) at target radii $0.99
R$ (left column) and $0.95 R$ (right column). From the top, the three
rows show reconstructions with RLS using GONG data, with RLS using MDI
data, and with OLA using MDI data.
\label{fig:F2}
}
\vspace{5mm}

At the highest latitudes, Figures~\ref{fig:F1} and \ref{fig:F2} show
more dynamical variations than at lower latitudes.  This is likely
related to the lesser moment of inertia associated with the polar
regions.  There may be evidence for the formation of a new band of
rapid rotation tending towards lower latitudes; the further evolution
of this feature should be followed over the coming years.

%
{\center\epsfig{file=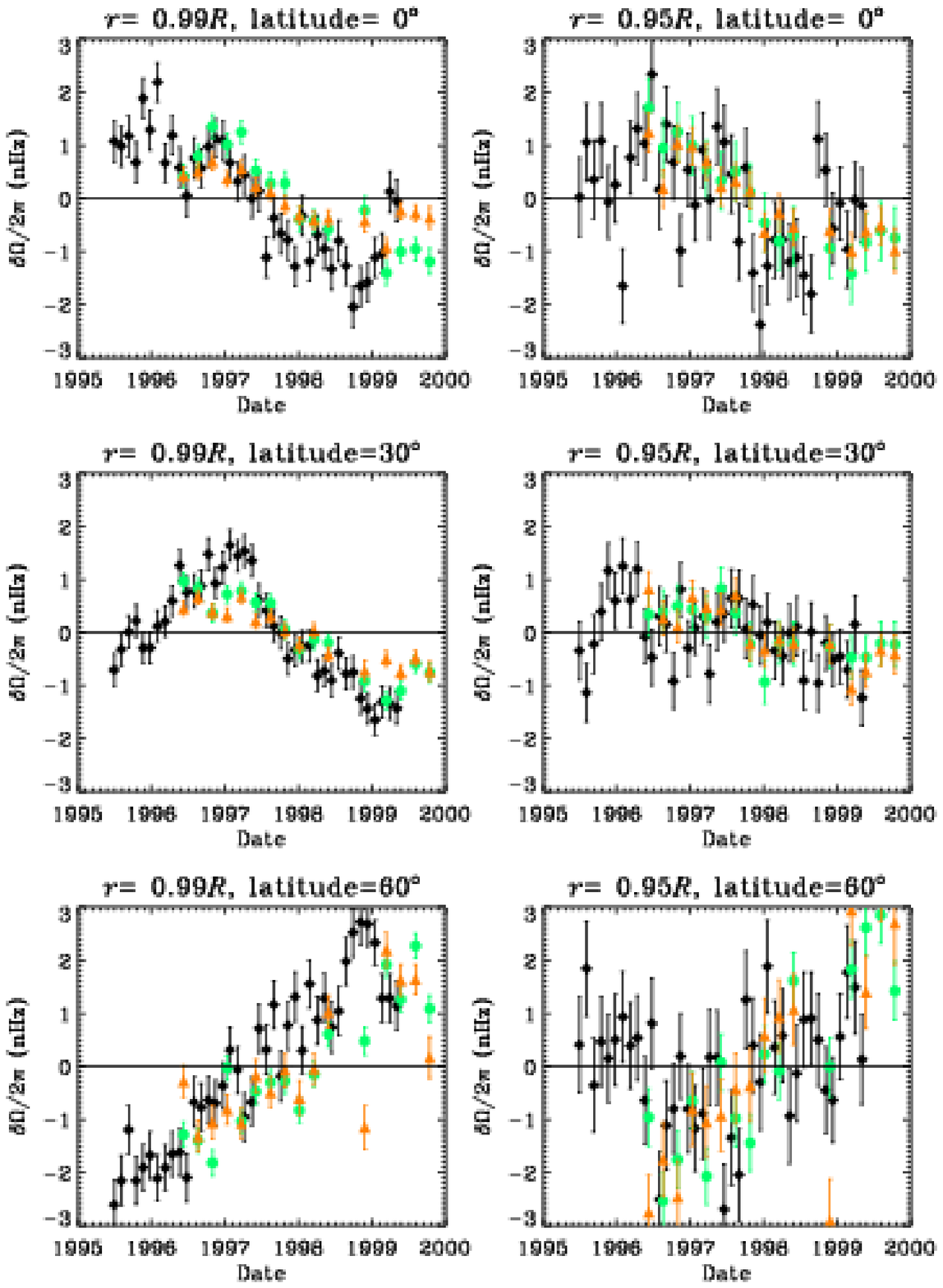,width=\linewidth}}
\figcaption{
Evolution with time in the residual rotation rate at target radii $0.99
R$ and $0.95 R$ and latitudes $0^\circ$, $30^\circ$, and $60^\circ$.
Diamonds (black) show results of RLS inversion applied to GONG data,
whereas circles (green) and triangles (orange) are for RLS and OLA
inversions on MDI data. The error bars show one standard deviation of
the inference, as determined from the errors in the observed data.
\label{fig:F3}
}
\vspace{5mm}

\placefigure{fig:F3}

\section{Conclusions}

Analysis of extended series of GONG and MDI data has revealed coherent
banded flow structures in the solar convection zone.  These correspond
to the torsional oscillations detected in direct Doppler observations
of the solar surface.  We have demonstrated that the flows are likely
to extend to a depth of at least 60 Mm, a substantial fraction of the
total 200 Mm depth of the convection zone, and considerably more than
the depth (about 35 Mm) at which the rotation rate attains its maximum
in the subsurface radial shear layer at low latitudes (cf. Schou et al.
1998a).  In addition,  there appear to be other systematic variations
with time of the residual rotation rate, with different signatures at
low and high latitudes (cf. Fig.~\ref{fig:F3}).

Inversions of global oscillation frequency splittings sample the
component of rotation symmetric around the equator.  The actual flows
will exhibit some level of asymmetry which will depend on the time
scale used in the analysis.  Indeed, local analyses by means of the
time-distance and ring-diagram techniques (Giles, Duvall, \& Scherrer
1998; Haber et al. 2000) have shown features similar to those found
here, but with clear differences between the two hemispheres.

%

{\center\epsfig{file=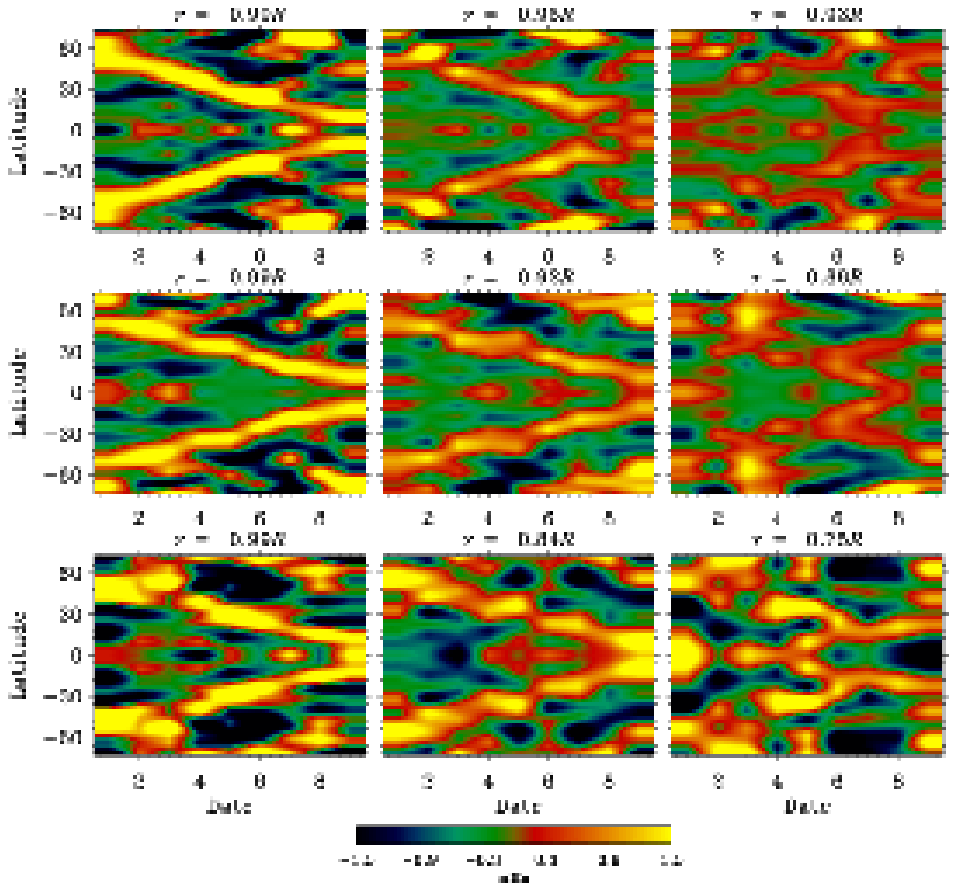,width=\linewidth}}
\figcaption{
Sample artificial data tests to study residual rotation rate inferred
from OLA inversions at various selected radii.  The imposed flow used
in the calculation of the frequency splittings extends from the surface
to $0.95 R$ in the top row, to $0.92 R$ in the middle row, and to $0.80
R$ in the bottom row.
\label{fig:F4}
}
\vspace{5mm} 

\placefigure{fig:F4}

The link between the evolving latitudinal positioning of the faster
zonal bands and of the sites of sunspot emergence suggest that the
dynamics are related, yet how this is accomplished is uncertain.  The
strong magnetic fields most likely originate from deep within the Sun,
probably formed by dynamo action near the base of the convection zone
(e.g., Spiegel \& Zahn 1992; Parker 1993; Weiss 1994; Charbonneau \&
MacGregor 1997).  Field bundles ascending from this region through the
convection zone, before erupting into the atmosphere as large-scale
magnetic loops, could well lead to significant perturbations in
velocity and thermal fields there.  This is likely to be accompanied by
some redistribution of angular momentum, given that the magnetic
structures will attempt to conserve their original angular momentum
(e.g., Brummell, Cattaneo \& Toomre 1995).  The coupling of a highly
turbulent medium with ascending magnetic structures, and their mutual
feedbacks, have not yet been assessed in recent flux-tube models.
Global simulations of turbulent convection in rotating spherical shells
(e.g., Elliott et al. 2000; Miesch et al. 2000; Miesch 2000) to study
the resulting differential rotation have revealed intrinsic variability
in zonal flows over intervals of several rotation periods, some of
which may be inertial oscillations (e.g., Gunther \& Gilman 1985), but
such modelling has not included large-scale magnetic fields.  Obtaining
propagating bands and time scales of variation of order the solar cycle
seems problematic unless there is some selective coupling to magnetic
processes.  Adding to the puzzle is that the evolving zonal bands are
present at the higher latitudes even before the prominent large-scale
magnetic eruptions begin (e.g. Ulrich 1998), as within this cycle.
Continued helioseismic observations as this magnetic cycle is
proceeding may help to provide clues about such aspects of solar
internal dynamics, for we now have the ability to probe hitherto unseen
flows well below the solar surface.

\section*{Acknowledgements}

This work utilizes data obtained by the GONG project, managed by the
National Solar Observatory, a Division of the National Optical
Astronomy Observatories, which is operated by AURA, Inc. under a
cooperative agreement with NSF. The data were acquired by instruments
operated by the Big Bear Solar Observatory, High Altitude Observatory,
Learmonth Solar Observatory, Udaipur Solar Observatory, Instituto de
Astrof\'{\i}sico de Canarias, and Cerro Tololo Interamerican
Observatory.  The Solar Oscillations Investigation (SOI) involving MDI
is supported by NASA grant NAG 5-3077 to Stanford University.  SOHO is
a mission of international cooperation between ESA and NASA.  RWK, and
RH in part, were supported by NASA contract S-92698-F.  JC-D was
supported by the Danish National Research Foundation through the
establishment of the Theoretical Astrophysics Center.  MJT was
supported in part by the UK Particle Physics \& Astronomy Research
Council.  JT was supported in part by NASA through grants NAG 5-7996
and NAG 5-8133, and by NSF through grant ATM-9731676.


\end{document}